\pdfoutput=1
\documentclass[sigconf]{acmart}

\usepackage{booktabs} 
\usepackage{tabularx}

\copyrightyear{2025}
\acmYear{2025}
\setcopyright{cc}
\setcctype[4.0]{by}
\acmConference[SAC '25]{The 40th ACM/SIGAPP Symposium on Applied Computing}{March 31-April 4, 2025}{Catania, Italy}
\acmBooktitle{The 40th ACM/SIGAPP Symposium on Applied Computing (SAC '25), March 31-April 4, 2025, Catania, Italy}
\acmDOI{10.1145/3672608.3707833}
\acmISBN{979-8-4007-0629-5/25/03}

\begin{document}
\title{Automated Market Makers:\\Toward More Profitable Liquidity Provisioning Strategies}
  
\renewcommand{\shorttitle}{Toward More Profitable Liquidity Provisioning Strategies}

\author{Thanos Drossos}
\authornote{Author to whom correspondence should be addressed.}
\orcid{0009-0001-6545-9096}
\affiliation{%
  \institution{Karlsruhe Institute of Technology}
  \streetaddress{Kaiserstr. 12}
  \city{Karlsruhe} 
  \country{Germany} 
  \postcode{76131}
}
\email{thanos.drossos@student.kit.edu}

\author{Daniel Kirste}
\orcid{0009-0007-3752-5109}
\affiliation{%
  \institution{Technical University of Munich}
  \city{Campus Heilbronn} 
  \country{Germany} 
  }
\email{daniel.kirste@tum.de}

\author{Niclas Kannengießer}
\orcid{0000-0002-2880-3361}
\affiliation{%
  \institution{Karlsruhe Institute of Technology}
  \streetaddress{Kaiserstr. 12}
  \city{Karlsruhe} 
  \country{Germany} 
  \postcode{76131}
  }
\email{niclas.kannengiesser@kit.edu}

\author{Ali Sunyaev}
\orcid{0000-0002-4353-8519}
\affiliation{%
  \institution{Technical University of Munich}
  \city{Campus Heilbronn} 
  \country{Germany} 
  }
\email{ali.sunyaev@tum.de}

\begin{abstract}
To trade tokens in cryptoeconomic systems, automated market makers (AMMs) typically rely on liquidity providers (LPs) that deposit tokens in exchange for rewards.
To profit from such rewards, LPs must use effective liquidity provisioning strategies. However, LPs lack guidance for developing such strategies, which often leads them to financial losses.
We developed a measurement model based on impermanent loss to analyze the influences of key parameters (i.e., liquidity pool type, position duration, position range size, and position size) of liquidity provisioning strategies on LPs' returns.
To reveal the influences of those key parameters on LPs' profits, we used the measurement model to analyze 700 days of historical liquidity provision data of Uniswap~v3.
By uncovering the influences of key parameters of liquidity provisioning strategies on profitability, this work supports LPs in developing more profitable strategies.
\end{abstract}



\maketitle

\section{Introduction}
\label{intro}

With the rise of decentralized finance, automated market makers (AMMs) became focal software agents that continuously offer token trades to market participants~\cite{Xu2023, Pourpouneh2020, Kirste2023}.
The AMM software code is commonly deployed to blockchain systems in the form of smart contracts.
Based on mathematical functions manifested in smart contract code, AMMs set token prices and automatically execute token trades~\cite{Kirste2023}.

To facilitate token trading, liquidity provider (LP)--based AMMs (e.g.,~Curve, Uniswap~v3, and Sushiswap~\cite{Adams, Xu2023}) rely on market participants acting as LPs by depositing tokens into designated liquidity pools~\cite{Adams, kirste_influence_2024}.
AMMs motivate market participants to become LPs by offering tokens as rewards in return for depositing tokens~\cite{Xu2023, Kirste2023}. Such rewards are a share of the trading fees charged by AMMs.

To increase rewards and thus profits, LPs use different strategies for managing tokens deposited in liquidity pools (referred to as positions in the following). Such strategies differ in multiple parameters, including position duration, position size, and position range size~\cite{Caparros2024}.
Although striving for higher rewards, LPs have mostly used strategies that resulted in very low or even no profits \cite{fritsch2024measuring} or financial losses \cite{Loesch2021}. Such losses are mostly due to realized impermanent loss (IL), a loss in value of the liquidity position compared to holding them outside.

One reason for LPs using such unprofitable strategies is that the influences of strategies' parameters on LPs' returns (rewards and realized IL) are unclear.
Consequently, LPs lack guidance on developing profitable liquidity provisioning strategies.
The influences of key parameters of liquidity provisioning strategies on profitability must be better understood to guide the development of more profitable strategies.
We ask the following research question:
\textit{How do key parameters of liquidity provisioning strategies influence LPs' returns?}

Using the concept of adverse selection costs from market microstructures~\cite{Aoyagi2020, Madhavan2000, Akerlof1970}, we devised a measurement model that can be used to compute the returns of LPs in constant-product AMMs with concentrated liquidity, such as Uniswap~v3~\cite{Adams}.
The measurement model uses the metric \textit{loss-versus-holding} (LVH) \cite{Aigner2021, Hashemseresht2022, Hafner, Heimbach} to compute IL and relate it to the fees gained by AMMs.
Next, we used the measurement model to analyze historical data from nine liquidity pools on Uniswap~v3 to reveal influences of key parameters of liquidity provisioning strategies on the profitability of liquidity provisioning. 

The main purpose of this study is to support the development of more profitable liquidity provisioning strategies for LPs in LP-based AMMs.
In particular, this work has the following contributions.
First, by presenting a measurement model to analyze the influences of key parameters of liquidity provisioning strategies on the returns of LPs, we enable analyses of the profitability of liquidity provisioning in AMMs.
Second, by analyzing the impact of these key parameters, we contribute to a better understanding of the influences of key parameters on liquidity provisioning on IL, rewards, and LP returns.
Third, by presenting four liquidity provisioning strategies, we offer actionable guidance to LPs to enhance profitability in liquidity provisioning. 



\section{Background and Related Research}

After briefly elucidating foundations of AMMs and IL important in this work, this section introduces key parameters in liquidity provisioning strategies. Subsequently, we describe related research and point out the need for this study.

\subsection{Automated Market Makers and Impermanent Loss}
\label{sec:background}

AMMs are software-based market makers that continuously quote bid (buy) and ask (sell) prices~\cite{Xu2023, Kirste2023}. This enables market participants to exchange tokens with the AMM at any time~\cite{Mohan2022, Fritscha}
Instead of order books, AMMs commonly employ mathematical pricing functions \cite{Kirste2023, Fritsch} implemented in the software code of smart contracts. Because the smart contract code is visible to users with access to a blockchain system \cite{kannengiesser2021smartcontract}, pricing functions are transparent.

To provide liquidity to markets, AMMs require LPs to deposit tokens in liquidity pools.
In return, LPs receive rewards in the form of tokens, which correspond to a share of the trading fees charged by an AMM for market making.
The amount of rewards results from the trading fees collected (typically $0.3$\%) and the proportion of an LP's liquidity relative to the total liquidity used to settle trades~\cite{Xu2023}.

Uniswap~v2 spreads liquidity across token prices $0$ to $\infty$, leading capital inefficiency since most of this liquidity remains unused, especially for correlated token pairs. Liquidity concentration functionalities, as in Uniswap v3, enable LPs to increase returns~\cite{Adams, Milionis2022} by concentrating tokens within narrower price ranges to earn more fees~\cite{Adams}, as they provide the same liquidity depth with less capital.

While profits through liquidity provisioning are possible, market makers, including AMMs, face an adverse selection problem~\cite{Akerlof1970}. This problem arises from information asymmetries between market participants and AMMs: market participants, who are better informed about external token prices, exploit stale AMM prices and thus buy underpriced (or sell overpriced) tokens from (or to) an AMM. Although this updates AMM prices, it also incurs financial losses for AMMs and their LPs~\cite{Madhavan2000, Glosten1985, Kyle1985}.

Adverse selection cost of AMMs can be quantified using the concept of impermanent loss (IL) \cite{Krishnamachari2021, Labadie2022}. IL refers to an LP's temporary monetary losses when providing liquidity compared to holding the tokens outright \cite{Angeris2021}.
IL increases when the token price ratio in a liquidity pool diverges from the price ratio at the time of deposit. IL is ``impermanent'' because it decreases when token prices revert to the price ratio at the time of deposit~\cite{Labadie2022}. However, when LPs withdraw positions, IL translates to permanent loss: LPs receive fewer of the higher-valued tokens and more of the lower-valued tokens~\cite{Milionis2023b, Xu2023}.
To gain profits from liquidity provisioning, LPs must use liquidity provisioning strategies that lead rewards to exceed realized IL.

\subsection{Liquidity Provisioning Strategies}

Liquidity provisioning strategies describe how LPs open or close positions to increase returns or limit losses.
Returns of an LP refer to the aggregate of rewards collected and realized IL.
Market participants execute different liquidity provisioning strategies that can include one or more positions.
The following describes key parameters (see Table~\ref{tab:parameters}) of liquidity provisioning strategies.

\paragraph{Pool Type}
Liquidity pools can contain stable and risky tokens. Stable tokens refer to currency-pegged stablecoins, such as USDC, USDT, and DAI. Risky tokens include the remaining cryptocurrencies, such as ETH, WBTC, and MKR. Risky tokens have higher price volatility than currency-pegged stablecoins.  
Based on the tokens in a liquidity pool, we differentiate three liquidity pool types: \textit{stable-stable}, \textit{stable-risky}, and \textit{risky-risky}. The token composition of the liquidity pools strongly influences the trading fees charged by the AMM and thus the rewards of the LPs~\cite{Heimbach2021}. 

\paragraph{Position Size}
The position size refers to the amount of tokens an LP deposits into an AMM's liquidity pool. To compare position sizes across liquidity pools, we denominate the position size $V^{LP}$ in a reference currency--USD in this work.

\paragraph{Position Range Size}
LP-based liquidity-concentrating AMMs (e.g., Uniswap~v3) enable LPs to increase rewards by specifying price ranges wherein their positions are used for market making.
The position range size $r$ only applies to LP-based liquidity-concentrating AMMs and specifies the price range of the positions relative to the current price at which LPs provide liquidity.


\paragraph{Position Duration}
Position duration refers to the time between opening and closing a position at an AMM's liquidity pool.
IL can be reduced by carefully timing the closure of a liquidity position. Thus, the duration of the position strongly influences the risks and returns of the positions \cite{Heimbach, Loesch2021}.


\begin{table}[b]
    \caption{Key parameters of liquidity provisioning strategies}
    \centering
    \begin{tabularx}{\linewidth}{p{1.6cm}|X}
        \hline
        \textbf{Parameter} & \textbf{Definition} \\
        \hline
        Pool Type & Classification of token risk and fee tier \\
        \hline
        Position \newline Duration & Time between opening and closing a liquidity position \\
        \hline
        Position Range Size & Relative price range within which tokens of a position can be used for liquidity provisioning \\
        \hline
        Position Size & Absolute USD value of tokens in a position  \\
        \hline
    \end{tabularx}
    \label{tab:parameters}
\end{table}

\subsection{Related Research on Liquidity Provisioning Strategies}
\label{sec:related-work}

Research on liquidity provisioning strategies revealed that, on average, LPs make losses compared to holding their tokens outside of liquidity pools~\cite{Loesch2021}. 
The average returns of LPs are usually low or even negative~\cite{Heimbach, Loesch2021}.
The returns of LPs vary between different types of pools~\cite{Heimbach2021}. Providing liquidity in stable-stable pools is a low-risk and marginally profitable investment. Stable-risky or risky-risky pools have the lowest returns due to high volatility that causes higher IL for LPs. 

If properly chosen, narrow position ranges can help increase returns compared to wide ranges due to capital concentration \cite{Caparros2024}.
However, narrow position range sizes increase the risks of negative returns in times of high volatility \cite{Heimbach}.
Narrow position ranges entail frequent adjustment of position ranges to center around the current token price. If the token price stated by the AMM exceeds a defined price range, the respective LP suffers impermanent (or even permanent) loss (if tokens are withdrawn from the liquidity pool).
Frequent adjustment of position ranges increases the cost of liquidity provisioning for LPs if a transaction processing fee is charged for each adjustment (e.g.,~gas in the Ethereum system).
Wide position range sizes help save gas fees because less repositioning is needed. In addition, setting wide price ranges close to the market price can mitigate IL.
Frequent adjustment generates more fees for LPs (but also gas fees), and higher liquidity concentration reduces slippage. This benefits both LPs and traders.

\citet{fritsch2024measuring} used the Loss-Versus-Rebalancing (LVR) to quantify arbitrage losses caused by stale prices in Uniswap~v2 and v3 liquidity pools.
The study shows that the liquidity pools with function-based liquidity concentration in Uniswap~v2 are more profitable than their corresponding Uniswap~v3 liquidity pools.
\citet{Cartea2022} revealed that frequently adjusting positions around the current token price in a liquidity pool can help LPs tackle IL and optimize their returns.
In complex liquidity provisioning strategies, high granular liquidity concentration and dynamic liquidity provision can provide significant gains for LPs~\cite{Fan2022}.

Although previous studies make valuable contributions to better understanding the influences of liquidity provisioning strategies on returns, they mainly focus on individual parameters such as position duration. The influences of other (combined) parameters, such as position size and position range size, on the returns of the LPs remain unclear.
To gain a deeper insight into how liquidity provisioning strategies affect LPs' returns, it is essential to thoroughly comprehend the impact of the critical parameters that entirely define these strategies on LPs' returns.
\section{Methods}
\label{methods}

We analyzed the influences of key parameters of liquidity provisioning strategies on LPs' returns in three steps, as described below.

\subsection{A Formalization of Impermanent Loss}

To quantify losses of LPs, we formalized IL as a foundation for analyzing the profitability of liquidity provisioning strategies~\cite{Milionis2022, Aigner2021, Loesch2021}.
Extant research presents two principal metrics to quantify IL: \textit{loss-versus-holding} (LVH) and \textit{loss-versus-rebalancing} (LVR). When using LVH, IL is calculated by comparing the value of the liquidity position with the corresponding portfolio following a buy-and-hold strategy based on the initial liquidity position~\cite{Aigner2021}.

Another approach to measuring IL uses LVR~\cite{Milionis2023a}, which calculates IL based on rebalancing strategies that hedge against adverse selection risks. Although LVR improves precision by isolating market risk and capturing adverse selection costs, its reliance on rebalancing makes it impractical for most investors and computationally complex, because every LP transaction must be replicated \cite{Milionis2023a}.
In contrast, LVH represents the practical decisions faced by investors, such as choosing between investing in a liquidity pool or simply holding tokens. On the other hand, LVR assumes a hypothetical, idealized scenario where liquidity providers continuously rebalance their position to maintain a specific token ratio, rather than just holding the tokens.
Therefore, we use the LVH equation~\ref{IL_v3} to measure IL in Uniswap~v3. In the following, a formalization of LVH will be developed.

AMMs with constant product price discovery mechanisms use a conservation function to determine token prices. The conservation function maps the token reserves amounts $x$ and $y$ to a constant $k$, which can be expressed as~\cite{Adams}:

\begin{equation}\label{cp_curve}
    x*y=k
\end{equation}

For AMMs with constant product and price discovery, IL increases with the progressing imbalance of the token reserve (inventory imbalance), which is needed by constant product price discovery to adjust token prices~\cite{Angeris2022When}.
IL refers to the unrealized loss that LPs experience compared to holding the tokens outside the liquidity pool. To measure IL, the value $V$ of tokens deposited in a liquidity pool at time $t$ must be determined.
The token value $V_t^{pool}$ can be computed as follows:

\begin{equation}
    V_t^{pool} = 2 * \sqrt{k*p_y^t*p_x^t}
\end{equation}
with $p_y^t$ and $p_x^t$ being the actual token prices at $t$.

The buy-and-hold (HODL) value $V_t^{HODL}$ is the value of the portfolio at time $t$ with a buy-and-hold strategy that does not participate in liquidity provisioning.

\begin{equation}
    V_t^{HODL} = x*p_x^t + y* p_y^t
\end{equation}

We define LVH as the relative return of tokens in the liquidity pool compared to a portfolio following a buy-and-hold strategy with identical initial tokens, taking the HODL value as a baseline:

\begin{equation}\label{LVH_normal}
    LVH(t) = \frac{V^{pool}_t - V^{HODL}_t}{V^{HODL}_t} = \frac{V^{pool}_t}{V^{HODL}_t} -1
\end{equation}

Equation~\ref{LVH_normal} can be rewritten as an expression for IL in constant-product AMMs, with $d$ denoting the relative price change or quotient of the token prices over time~\cite{Pourpouneh2020}:

\begin{equation}\label{IL_normal}
    LVH_{v2}(d) = \frac{2*\sqrt{d}}{1+d} -1
\end{equation}

This equation can be adapted to AMMs with concentrated liquidity with the lower price bound $p_a$ and upper price bound $p_b$~\cite{Hashemseresht2022}:

\begin{equation}\label{IL_v3}
IL_{v3} =
\begin{cases}
\frac{\sqrt{\frac{p_b}{p}} - d \left(1 - \sqrt{\frac{p}{p_b}}\right) - 1}{d \left(1 - \sqrt{\frac{p}{p_b}}\right) + 1 - \sqrt{\frac{p_a}{p}}}, & \text{if } \frac{p_b}{p} < d \\
\frac{2 \sqrt{d} - 1 - d}{1 + d - \sqrt{\frac{p_a}{p}} - d \left(\sqrt{\frac{p}{p_b}}\right)}, & \text{if } \frac{p_a}{p} \leq d \leq \frac{p_b}{p} \\
\frac{d \left(\sqrt{\frac{p}{p_a}} - 1\right) - 1 + \sqrt{\frac{p_a}{p}}}{d \left(1 - \sqrt{\frac{p}{p_b}}\right) + 1 - \sqrt{\frac{p_a}{p}}}, & \text{if } d < \frac{p_a}{p}
\end{cases}
\end{equation}

\newpage

\subsection{Data Gathering and Preparation}

To prepare a data set for the subsequent analysis, we collected historical liquidity data from Uniswap~v3 using the Uniswap~v3 subgraphs\footnote{Revert finance subgraph: \url{https://api.thegraph.com/subgraphs/name/revert-finance/uniswap-v3-mainnet/}\\Uniswap subgraph: \url{https://api.thegraph.com/subgraphs/name/uniswap/uniswap-v3/}}. The choice to analyze Uniswap v3 is supported by its largest DEX market share of 15\% in April 2024, making it a dominant representative of LP-based AMMs \cite{CoinMarketCap}. 
The data set included position sizes (deposited and withdrawn position values in USD and quantities), dates (deposit and withdrawal time), pool types (tokens and fee tier), position ranges (upper and lower price ticks), fee collection events, and general liquidity pool data (volume, total value locked, and transaction count). 
Moreover, the data set contained records of all LP transactions with nine liquidity pools of Uniswap~v3 (see Table~\ref{table:pools}). The selected pools represent 16\% of the total value locked on Uniswap v3 during the analysis period.
Uniswap v3 uses liquidity pools with different fee tiers for the same trading pairs.
Some liquidity pools offer high liquidity and are commonly used, while others suffer from low liquidity, which decreases the efficiency of token exchanges and makes them less frequently utilized.
To decrease bias from too low liquidity in the following analysis, we removed data related to pool types with less than USD 10,000 in total value locked (see Table~\ref{table:pools}).
We used the Uniswap liquidity pool naming convention, containing the liquidity pool tokens and the fee tier in 100 base points (e.g.,~DAI-USDC-500 for the DAI-USDC 0.05~\% fee liquidity pool).
As stable-stable liquidity pools typically hardly incur IL, we used them as a baseline for comparison with other liquidity pools regarding LP profits.

We prepared the final data set for the analysis by extracting LP transactions over a period of 700 days, starting from May 1, 2022 (block 14,691,320) until April 1, 2024 (block 19,560,244). We only included positions in the final data set that were closed during that time. 
In that time, several notable events occurred, including sharp market downturns, such as the collapse of FTX in November 2022, strong market upturns, such as the Ethereum Merge in September 2022, and sideways movement with a slight overall uptrend driven by institutional adoption of cryptoeconomic systems~\cite{Caparros2024, Babalos2024}.

\begin{table}[h]
    \caption{Analyzed liquidity pools of Uniswap~v3. The pool types excluded from the analysis are marked with a dash (-).}
    \label{table:pools}
    \centering
        \begin{tabular}{l@{\hspace{3pt}}|@{\hspace{3pt}}c@{\hspace{3pt}}c@{\hspace{3pt}}c@{\hspace{3pt}}c}
            \hline
            \textbf{Type\textbackslash Fee} & \textbf{1\%} & \textbf{0.3\%} & \textbf{0.05\%} & \textbf{0.01\%} \\
            \hline
            Stable-Stable & - & - & DAI-USDC & DAI-USDC \\
            Stable-Risky & USDC-ETH & USDC-ETH & USDC-ETH & USDC-ETH\\
            Risky-Risky & MKR-ETH & BTC-ETH & BTC-ETH & - \\
            \hline
        \end{tabular}
\end{table}

\subsection{Data Analysis}

To analyze the influences of key parameters on LPs' returns, we analyzed liquidity positions by measuring IL, rewards, and LP returns.
We calculated the realized IL by matching deposit and withdrawal events, with liquidity additions/removals handled using a first-in-first-out approach.
Then, we conducted a two-step analysis on the individual influences of each key parameter on IL, rewards, and LPs' returns isolated.
First, we analyzed IL and rewards for all pool types to create a baseline. We used this baseline as a reference to compare influences of key parameters in liquidity provisioning strategies on LPs' returns.
Second, we analyzed IL and rewards for the remaining parameters across all pool types in cumulative metrics.
Next, we defined the liquidity provisioning strategies based on the position duration and position range size. We analyzed those strategies for different pool types (i.e.,  stable-risky, risky-risky) by calculating IL, rewards, and LP returns.
We excluded stable-stable pool types from the analysis of liquidity provisioning strategies because IL is commonly negligible for these pool types.
We added 95\% confidence intervals of the average returns to assess the robustness of our results.
We differentiated positions along the key parameters \textit{position duration} into short-term and long-term and \textit{position range size} into narrow-range and wide-range. We chose the cut-off thresholds as the 30th and 70th percentile to determine the lower and upper parameters, respectively.

\begin{table}[t]
    \caption{Parameter configurations for liquidity provisioning strategies, cutoff thresholds based on 30th/70th percentiles}
    \centering
    \begin{tabular}{l|@{\hspace{3pt}}c@{\hspace{3pt}}c}
        \hline
        \textbf{Liquidity Position Strategy} & \textbf{Duration} & \textbf{Range Size} \\
        \hline
        Short-Term Narrow-Range Strategy & $< 1.12$ days & $< 0.0467$ \\
        \hline
        Long-Term Narrow-Range Strategy & $> 26.90$ days & $< 0.0467$ \\
        \hline
        Short-Term Wide-Range Strategy & $< 1.12$ days & $> 0.2756$ \\
        \hline
        Long-Term Wide-Range Strategy & $> 26.90$ days & $> 0.2756$\\
        \hline
    \end{tabular}
    \label{tab:trader_types}
    
\vspace{-17pt}

\end{table}
\section{Results}

This section describes the influences of key parameters of liquidity provisioning strategies on IL, rewards, and LPs' returns.
In section~\ref{sec:individual-key-parameters}, we describe IL and rewards for all pool types to create a baseline. Using this baseline, we compare liquidity provisioning strategies regarding their profitability.
In section~\ref{sec:liquidity_strategies}, we describe the results from the analysis of the remaining parameters across all pool types.

\subsection{Analysis of Individual Key Parameters}
\label{sec:individual-key-parameters}

This subsection describes the influences of each key parameter on IL and rewards of LPs. Except for the pool type analysis, all results are aggregated over all non-stablecoin-only pools in our sample.

\subsubsection{Pool Types}
\begin{figure}[b]
    \centering
    \includegraphics[width=1\linewidth]{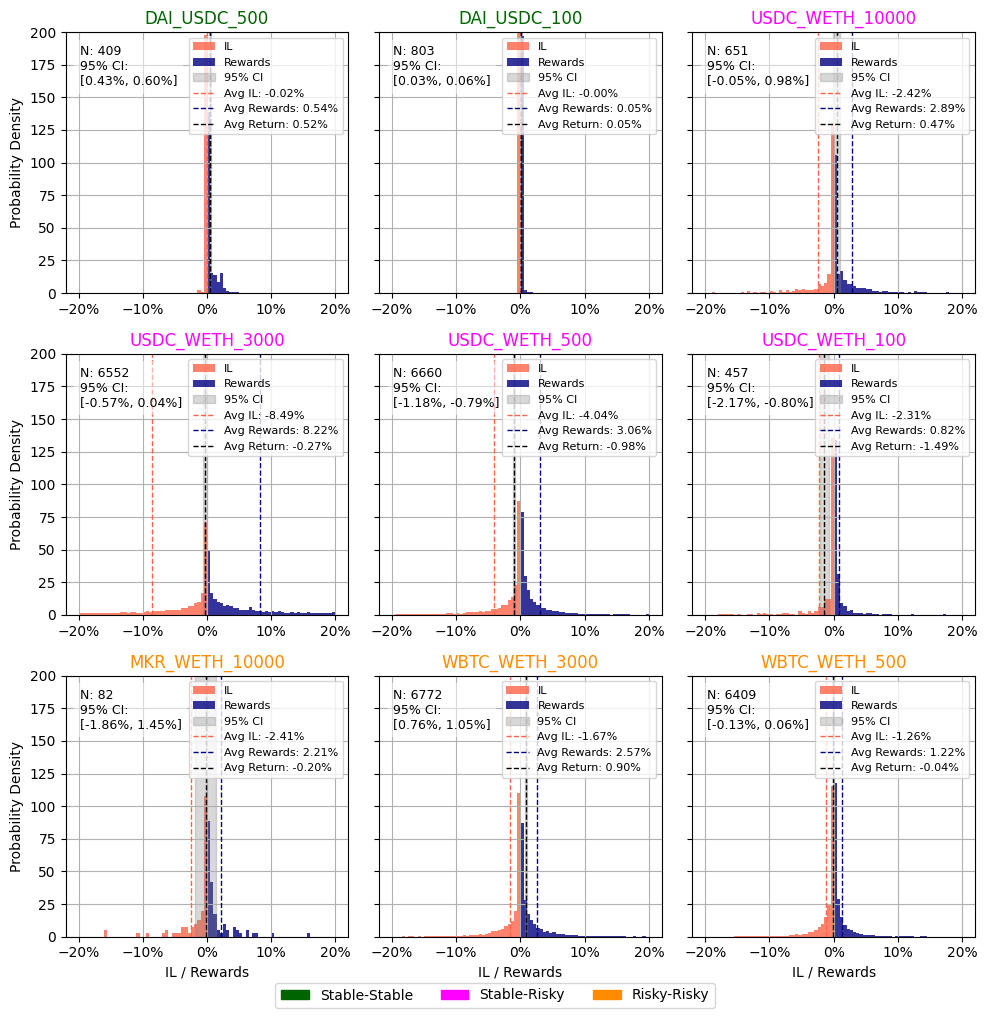}
    \caption{Distribution of IL (red) and rewards (blue) for all liquidity pools. Stable-stable pool types show negligible IL, while stable-risky and risky-risky pool types experience high IL. The fee tier strongly influences rewards. }
    \Description{A figure with histograms of the IL and fee distribution of the nine liquidity pools, showing that the fee tier strongly influences rewards. IL and fees are both centered at 0 and then decrease exponentially.}
    \label{fig:IL_Fees_pools}
\end{figure}

The histograms in Figure~\ref{fig:IL_Fees_pools} illustrate the distribution of IL and rewards across the nine analyzed liquidity pools. The distributions of IL and rewards are concentrated around zero and decrease exponentially.
The spike in the distribution of IL and rewards around zero can be attributed to two types of liquidity provisioning strategies: \textit{(1)} short-term liquidity provisioning, which generates low rewards and incurs low IL, and \textit{(2)} stable-stable pool type liquidity provisioning that leads to negligible IL because the liquidity pool's tokens are strongly correlated in token prices. In any case, the densities of IL and rewards decrease exponentially.

Furthermore, Figure~\ref{fig:IL_Fees_pools} illustrates the IL and the rewards of each liquidity pool. It can be observed that stable-risky pools exhibit higher IL than rewards.
LPs have negative returns of -0.61\% on average when providing liquidity in stable-risky pools (with highly uncorrelated tokens) and slightly positive returns for stable-stable (0.23\%) and risky-risky pools (0.44\%) .

\subsubsection{Position Duration}
\begin{figure}[h]
    \centering
    \includegraphics[width=1\linewidth]{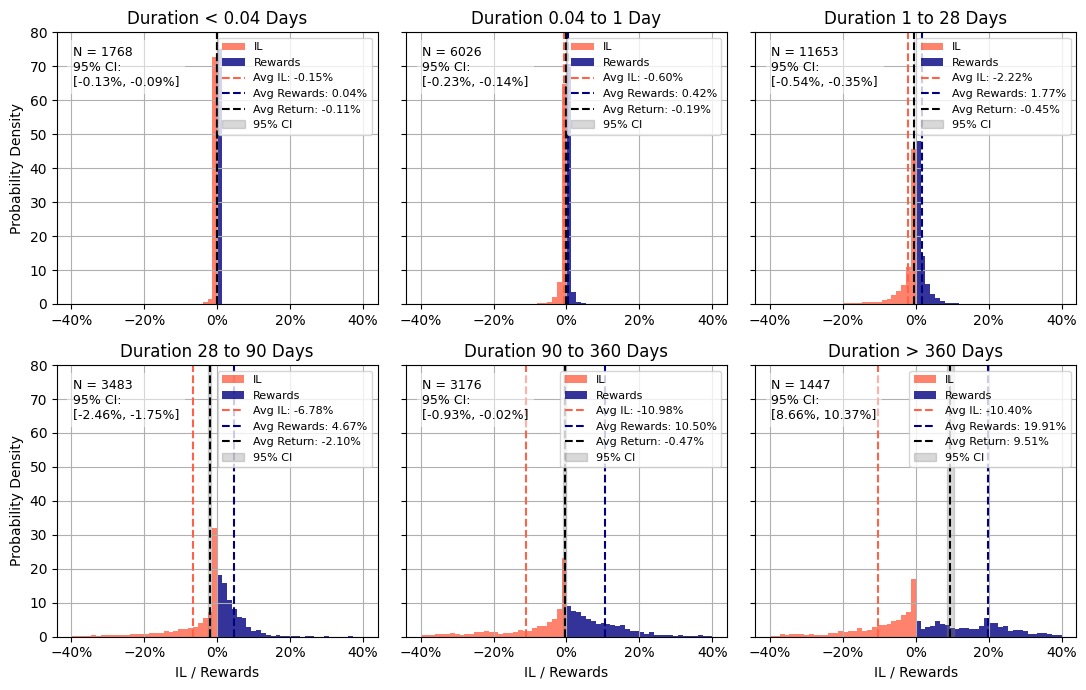}
    \caption{Rewards and IL by position duration. The data was split into six duration buckets and average IL and rewards are plotted in each histogram.}
    \Description{Six histograms of IL and rewards, showing that the distribution of IL and rewards flattens with increasing time.}
    \label{fig:comp_duration}
\end{figure}

Figure~\ref{fig:comp_duration} illustrates IL and rewards for different position durations. IL and rewards both increase with longer position durations. 

The division of IL and rewards by position duration shows that the earned rewards grow slightly faster than IL over time (see Table~\ref{tab:daily}).  
The only profitable positions seem to be long-term positions (more than 360 days), yielding around 19.9~\% in rewards but only recording an average IL of 10.4~\%.

Short-time positions (less than 1 hour) mostly lead to negative returns, where IL outweighs rewards. Short-term positions have a negligible median daily IL. For slightly longer durations, the daily IL rises and then gradually decreases to lower levels for positions over 28 days. Positions with a duration of between one hour and one day exhibit the highest daily IL and rewards. Similarly to mean values, median rewards grow faster than IL over time. IL is close to zero for very short-timed positions. 

Based on these observations, we conclude that \textit{positions with long position durations result in high IL and high rewards, while rewards only outweigh IL for position durations longer than 360 days.}

\begin{table}[h]
    \caption{Comparison of median (daily) IL and fees for different position durations.}
    \centering
    \begin{tabular}{c|>{\centering\arraybackslash}p{1.2cm}>{\centering\arraybackslash}p{1.2cm}>{\centering\arraybackslash}p{1.2cm}>{\centering\arraybackslash}p{1.2cm}}
         Duration &  Daily IL & Daily rewards& IL & rewards\\
         \toprule
         < 1 hour & $<~\text{-}0.001~\% $ & 0.04~\% & $<~\text{-}0.001~\% $& 0.008~\%\\
         1 hour to 1 day & -0.26~\% & 0.32~\% & -0.12~\% & 0.13~\%\\
         1-28 days & -0.13~\% & 0.14~\% & -0.71~\% & 0.82~\% \\
         28-90 days & -0.04~\% & 0.06~\% & -2.42~\% & 3.05~\%\\
         90-360 days & -0.03~\% & 0.04~\% & -4.90~\% & 7.64~\%\\
         360 + days & -0.01~\% & 0.03~\% & -6.20~\% & 18.13~\%\\
    \end{tabular}
    \label{tab:daily}
\end{table}

\subsubsection{Position Size}
\begin{figure}[b]
    \centering
    \includegraphics[width=1\linewidth]{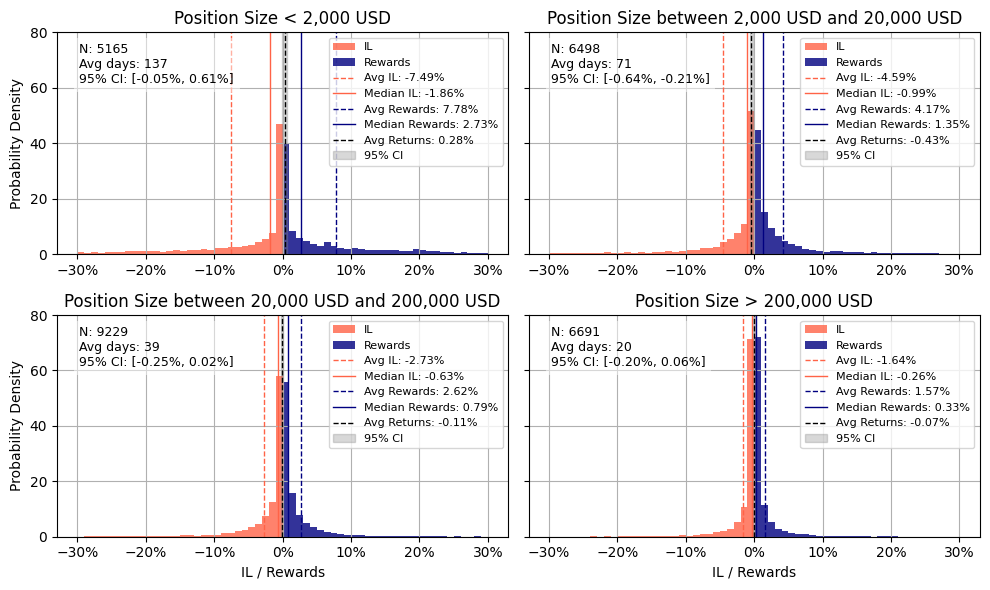}
    \caption{Rewards and IL by position size.}
    \Description{Four histograms of IL and rewards, showing that both decrease with increasing position sizes and are thus more concentrated around 0.}
    \label{fig:comp_size}
    \vspace{-6pt}
\end{figure}

Figure~\ref{fig:comp_size} illustrates the IL and rewards of positions grouped by position size. With increasing position size, IL and rewards decrease. The large difference between the mean and median of IL and rewards hints at outliers on both ends, especially for small positions. The mean return is minimally influenced by the position size: the smallest positions yield marginal gains of approximately 0.28~\%, while larger positions incur minor losses within the range of 0.07~\% to 0.43~\%. 

We assume that position duration might cause this decrease in IL and rewards for large position sizes because positions with large sizes typically have shorter durations than positions with small sizes.
We only observed a few long-term positions with a size larger than 100,000 USD. Large-volume traders seem to prefer short position durations of less than 20 days. 
The observed decrease in rewards and IL with increasing position size can be partially explained by the fact that larger positions are often associated with shorter durations. 
LPs seem to reduce the risks related to large-volume positions by many short-term positions.
In our analysis, position size has little influence on the profitability. The overall returns remain roughly the same for all position sizes.
For these reasons, position size is not included in the analysis of LP strategies.

Our observations still suggest that \textit{IL and rewards decrease with increasing position sizes.}

\subsubsection{Position Range Size}

\begin{figure}[b]
    \centering
    \includegraphics[width=1\linewidth]{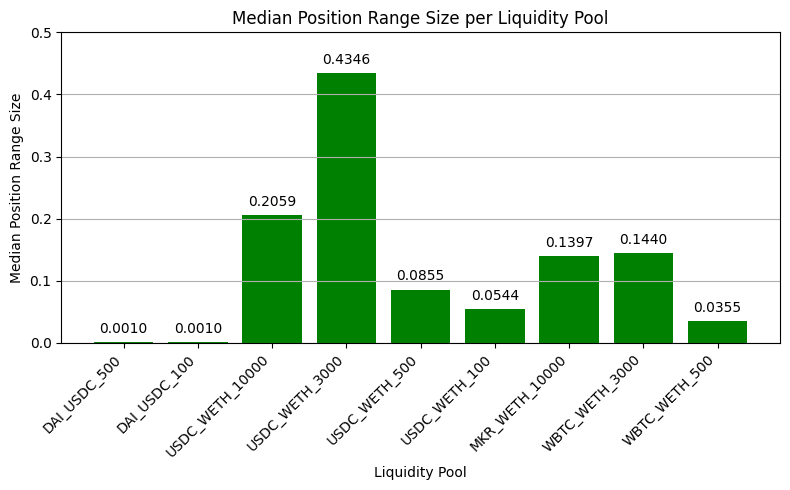}
    \caption{Median position range size of all liquidity pools.}
    \Description{Bar chart showing the median range size per liquidity pool. USDC-WETH-3000 has the highest position range size (0.43), followed by USDC-WETH-10000 (0.2). Stable-stable pools have very narrow range sizes of around 0.001.}
    \label{fig:range_dist}
    \vspace{-6pt}
\end{figure}

Figure \ref{fig:range_dist} illustrates the average position range size across all analyzed pools. Stable-stable pools show very small position range sizes, which can be attributed to commonly very low price fluctuations in such liquidity pools~\cite{Heimbach}. We identified the largest price ranges in stable-risky liquidity pools, which commonly show the largest price volatility. 



\begin{figure}[b]
    \centering
    \includegraphics[width=1\linewidth]{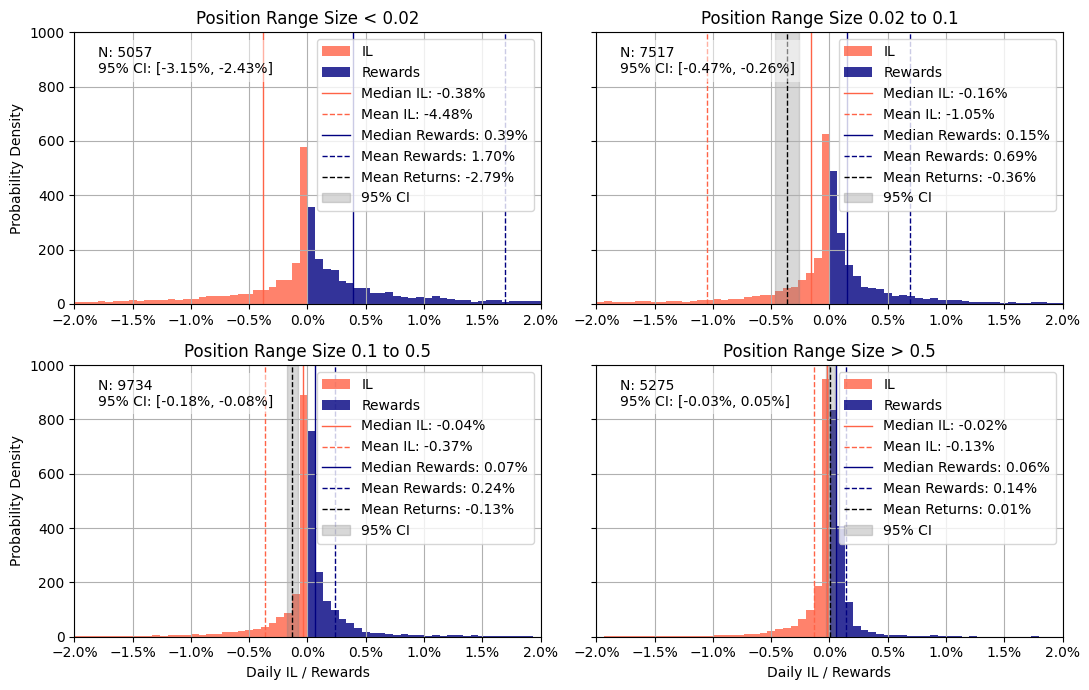}
    \caption[Daily rewards and IL by position range.]{Daily rewards and IL by position range.
    }
    \Description{Four histrograms of the IL and rewards distributions over different position range sizes, showing that larger range sizes lead to IL and fees being more concentrated around 0.}
    \label{fig:comp_range}
    \vspace{-6pt}
\end{figure}

Figure~\ref{fig:comp_range} illustrates the distribution of IL and rewards for different position range sizes.
Narrow range sizes incur a higher daily IL than wide range sizes. The relative daily rewards earned by a narrow price range are high because of better capital efficiency (concentrated liquidity). However, IL for positions with narrow range sizes outweighs the rewards. In particular, LPs with range sizes smaller than 0.02 experience a median IL of around 0.38~\% per day.
Only range sizes wider than 0.5 have (slightly) positive returns.

Based on these observations, we conclude that \textit{positions with larger position range sizes carry less IL and earn fewer rewards; however, they are the only positions with positive returns.}

\subsection{Liquidity Provisioning Strategy Analysis}
\label{sec:liquidity_strategies}

\begin{figure}[b]
    \centering
    \includegraphics[width=1\linewidth]{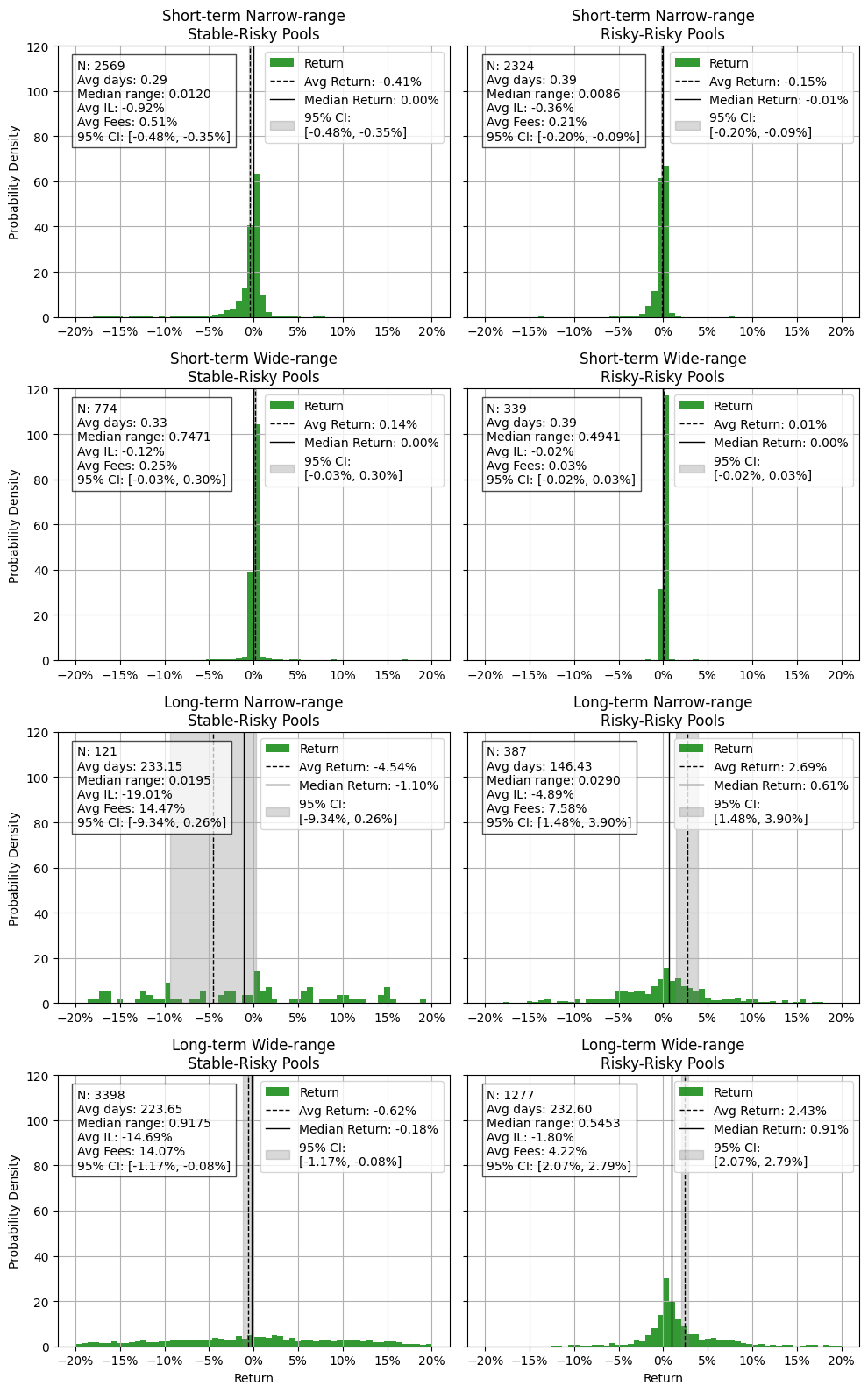}
    \caption{Returns of liquidity provisioning strategies for different pool types.}
    \Description{A figure showing eight histograms of the returns of each strategy for the pool types. Short-term wide-range positions have very little returns, while long-term wide-range positions have higher average returns of 2.4\%. Long-term wide-range position have returns almost uniformly distributed between -20\% and 20\%. Long-term narrow-range positions in stable risky pools lead to negative returns on average, with the returns being widely distributed.}
    \label{fig:comp_strategies}
    \vspace{-12pt}
\end{figure}

Figure \ref{fig:comp_strategies} presents the analysis of four distinct liquidity provisioning strategies that differ in position duration and position range size. These four strategies are analyzed in the context of two pool types (i.e., stable-risky and risky-risky).

We observe a negative correlation between the position range size and position duration. Narrow-range positions are often closed faster, resulting in shorter durations, than wide-range positions. This can be observed for a large number of positions (N in Figure~\ref{fig:comp_strategies}) for short-term narrow-range and long-term wide-range strategies (see first and fourth row in Figure~\ref{fig:comp_strategies}). In contrast, short-term wide-range and long-term narrow-range strategies are rarely represented in the analyzed dataset (see second and third row in Figure~\ref{fig:comp_strategies}).
We assume that LPs with narrow ranges are more sensitive to price fluctuations and close their positions sooner to limit potential IL. This leads to a form of survivorship bias in our sample: few narrow-size positions are likely to reach a long-term holding period~\cite{brown_survivorship}.

Short-term positions seem to hardly generate positive returns, especially when employing narrow-range liquidity provisioning strategies (see average return in the upper four subplots). An exception is observed for short-term wide-range strategies within stable-risky pools, where they can achieve positive returns of 0.14\%. This indicates that wide ranges may offer better protection against short-term volatility in these pools than low ranges. However, long-term strategies outperform short-term ones due to two primary factors: \textit{(1)} long position durations allow for accumulation of more rewards, and \textit{(2)} IL seems to decrease over time.

Liquidity provisioning strategies in risky-risky pools seem more profitable than those in stable-risky pools. This can be attributed to the observed positive correlation between tokens in risky-risky pools, which decreases volatility~\cite{Gkillas2018}.

The long-term narrow-range strategy yields the highest average returns of 2.69\% but also comes with a high risk, evidenced by a relatively large variance in the returns.
For stable-risky pools, the short-term wide-range strategy is the most profitable, with average returns of 0.14\%. Considering the average duration of 0.33 days in this strategy, such returns might become substantial if this strategy is repeated multiple times.

Selecting an appropriate position range size and duration for liquidity provisioning strategies is crucial and should align with the expected price drift of the pool tokens. In stable-risky pools, employing short-term, wide-range strategies is advantageous because of the high volatility of the token pair. In risky-risky pools, short-term strategies are generally not profitable. In long-term setting, narrow ranges capitalize on the correlated price movements of risky-risky tokens, with these positions being slightly more profitable than wide-ranged ones (due to concentrated liquidity). 
Finally, the analysis revealed that position range size influences returns less in risky-risky pools than in stable-risky pools. This is because the price drift between tokens in risky-risky pools is typically smaller than in stable-risky pools.

\section{Discussion}
\label{discussion}

This section presents the key findings, elucidates this work's contributions, and describes its limitations and future research directions.

\subsection{Principal Findings}


The distribution of IL across liquidity pools shows that most LPs use liquidity provisioning strategies to reduce the IL, achieving an IL of -3.8\% on average. We recognized that more than half of LPs have an IL of less than -0.8\%. However, we also observed that most LPs could not generate substantial profits because realized IL outweighed rewards. The overall returns were between -1\% and 1\% for more than 50\% of LPs. We also observed that 49.5\% of the positions generated negative returns from providing liquidity. Thus, average LPs neither profit nor lose large amounts of capital from providing liquidity but have close-to-zero returns.
This finding highlights the importance of suitable liquidity provisioning strategies to enable LPs to profit from liquidity provisioning.

Our analysis shows that the pool types (token risks and fee tier) influence IL and rewards. Stable-risky pools carry the highest IL and rewards, but negative returns on average. Pools with correlated tokens had positive returns on average, with risky-risky pools having the highest returns due to their high fee structure. LPs could only consistently yield positive returns from low IL and low rewards in stable-stable pools.
For pools with correlated tokens (e.g., risky-risky pools), long-term narrow-range strategies perform best. For pools with less correlated tokens (e.g., stable-risky), short-term wide-range strategies yield the highest returns.
Based on those findings, we conclude that decisions on the pool type to which LPs provide liquidity seem of great importance in liquidity provisioning strategies that yield positive returns. 

The results show large differences in IL between liquidity pools managing the same tokens but having different fee tiers. Except for the USDC-WETH-10000 liquidity pool, the average IL per liquidity pool increases with a growing fee tier.
We assume that liquidity pools charging high fees attract risk-seeking LPs who take riskier liquidity positions (e.g.,~by selecting smaller range sizes) to gain higher rewards.
The distributions of IL and rewards are centered around zero for all liquidity pools but expand when the expected price drift between the two tokens and the fee tier increases. The mean values of IL and rewards increase with the risk level of the liquidity pools. However, IL and rewards balance out, resulting in an average zero return.

Positions with narrow price ranges are exposed to high IL but can yield high rewards if managed actively and appropriately. However, our analysis shows that the risk of losses due to narrow ranges outweighs the possible higher rewards associated with narrow ranges in stable-risky pools.

\subsection{Contributions}

This work contributes to the field of AMMs in three principal ways.
First, we present a measurement model to analyze the influences of key parameters of liquidity provisioning strategies on LPs' returns to support better profitability evaluation in liquidity provisioning.

Second, this work supports a better understanding of how key parameters influence profitability in liquidity provisioning. Such understanding forms a foundation for actionable guidance for LPs in leveraging these parameters to increase profitability.

Third, we present four liquidity provisioning strategies based on position duration and range size for two liquidity pool types. By evaluating those strategies, we contribute to a better understanding of how realized IL and rewards influence the profitability of liquidity provisioning. This can help LPs to increase their returns.

\subsection{Limitations and Future Research}

The measurement model only accounts for realized IL, neglects gas fees, and does not consider the actual volatility of token pairs.
Gas fees would negatively impact the returns of smaller positions more than larger positions.
Implementing LVR~\cite{Milionis2022} as a metric for delta-hedged liquidity positions could help isolate market risk to provide a more accurate measurement of IL. Since our analysis is based only on historical data, it might not be entirely applicable to predicting future scenarios.

Furthermore, our sample might include survivorship bias since some unprofitable LPs might never have been closed (e.g., due to high IL with liquidity providers waiting to revert). This could be addressed with simulations of unrealized IL.

This work concentrates on Uniswap~v3, a popular representative of LP-based liquidity-concentrating AMMs. The results are only applicable to such AMM designs. Other AMM designs may exhibit other market conditions and results.
To better understand how the key parameters influence profitability in other settings, this work should be extended by analyses that incorporate other AMM types (e.g.,~price-adopting AMMs), examine more liquidity pools, and extend the analysis period to decrease influences of market events.

Further research should explore other profitability metrics to explain why LPs provide liquidity, despite incurring losses on average.

\section{Conclusion}

Liquidity provisioning in AMMs carries substantial risks for LPs, essentially financial losses caused by realized IL, which can shy away LPs and, thus, could lead to depletion of liquidity in AMMs.
IL can be compared to adverse selection costs caused by inventory imbalances in AMMs.
On average, realized IL led to a 3.8\% loss per position compared to a simple buy-and-hold strategy.
Our analysis shows that such losses can be mitigated by employing strategies like providing liquidity in pools with correlated tokens, using larger position range sizes, and longer position durations for liquidity provision. These strategies should suite the pool type and its expected token price volatility to be profitable.

While AMMs like Uniswap~v3 offer innovative mechanisms for decentralized token exchanges, the economic viability of liquidity provisioning remains constrained. The observed neutral to slightly negative returns emphasize the need to make liquidity provisioning more profitable to attract LPs and enhance the reliability and viability of cryptoeconomic systems.

\section*{Acknowledgment}
This work was supported by funding from the topic Engineering Secure Systems of the Helmholtz Association (HGF) and by KASTEL Security Research Labs.


\bibliographystyle{ACM-Reference-Format}
\bibliography{sample-bibliography} 

\end{document}